\documentclass[12pt,preprint]{aastex} 

\shorttitle{$^6$Li in the atmosphere of GJ~117}
\shortauthors{M. Mathioudakis et~al.}
\begin{document}

\title{$^6$Li in the atmosphere of GJ 117}

\author{D.J. Christian, M. Mathioudakis,}
\affil{Department of Physics and Astronomy, Queen's University Belfast,
	Belfast, BT7 1NN, Northern Ireland, U.K.}
\and
\author{ D. Jevremovi\'c }
\affil{Department of Physics and Astronomy, University of Oklahoma, 440 W. Brooks NH 131, Norman 73019, USA and Astronomical Observatory, Volgina 7, 11160 Belgrade, Serbia and Montenegro }
\and
\author{ P. H. Hauschildt }
\affil {Hamburger Sternwarte,  Gojenbergsweg  112, 21029  Hamburg, Germany}
\and
\author{E. Baron}
\affil{Department of Physics and Astronomy, University of Oklahoma, 440 W. Brooks NH 131, Norman 73019, USA }



\begin{abstract}
We present high resolution VLT UVES observations of the active K dwarf GJ~117. 
$^6$Li enhancement has been shown for energetic
solar events, one chromospherically active binary, and several dwarf
halo stars.
Our analysis reveals the detection of $^6$Li on this source with $\frac{^{6}Li}{^{7}Li}$ = 0.030$\pm0.010$. 
We found no significant contribution from other lines, including Ti I, 
in the Li profile of GJ 117 and a template star of similar spectral type and metallicity. 
We discuss the possibility for $^6$Li production by spallation 
and find it to be consistent with the activity levels of the object.  

\end{abstract}


\section{Introduction}
\label{intro}

Lithium is a fragile element that serves as a powerful diagnostic of
stellar structure and evolution. The fragility of Li arises from the
fact that it can burn at relatively low temperatures via the [$^7$Li
(p,$\alpha$) $^4$He] and [$^6$Li (p,$\alpha$) $^3$He] processes.  Most
of the Li destruction occurs during the pre-main sequence state, when
the convection zones are deepest and for stellar models with the
lowest mass.  Lower mass stars arrive on the ZAMS at lower
temperatures and the amount of protostellar Li that survives today in
dwarf stars decreases with decreasing temperature \citep{DM95}.  
Stellar clusters such as $\alpha$ Per and the
Pleiades, show a large spread of Li abundances at each effective
temperature and it has been shown that the largest Li abundances are
found in fast rotating stars with the strongest chromospheric emission
\citep{S93}. This is a puzzling result as theoretical
models predict that fast rotation enhances the mixing process that
leads to increased Li depletion; the rotational history plays a 
major role in the depletion process \citep{P97}. 

On the main sequence, the correlation between Li and chromospheric activity is
not very strong. Stars with the same Ca II flux may have Li abundances that
differ by one order of magnitude. Lithium is enhanced in sunspots \citep{G84} 
where a small
but notable amount of $^6$Li has been detected
\citep{R97}. An investigation of 
active RS CVn binaries failed to show a clear correlation between
photometric variations and Li abundances \citep{P92}. Plages and
active regions could play a minor role but their effect would be to decrease
rather than increase the Li 6707.8\AA\ equivalent width 
\citep{HD95}. \citet{Ru96} used the Li 6104\AA\ line, which is formed deeper 
into the
atmosphere, to study a sample of Pleiades K dwarfs. The results showed that the
derived abundances have significantly less scatter than those derived from the
6707.8\AA\ line. If activity has an effect on Li, it seems to increase rather
than decrease the equivalent width and is therefore opposite to what the
chromospheric NLTE calculations predict. 

Surface activity could increase the Li
abundance if extreme energetic conditions are met \citep{ML99}.  
Spallation reactions from low energy flares can increase 
$\frac{^{6}Li}{^{7}Li}$ ratios \citep{C75}.
Predictions for significant creation of 
$^6$Li in stellar and solar flares have been made
\citep{C75, W85, L97}  
although actual measurements for the Sun have found the 
average $\frac{^{6}Li}{^{7}Li}$ of 0.01 and up to 0.04 
\citep{TR71,M75}.

The first detection of $^6$Li on the halo dwarf HD 84937 has been highly 
significant 
\citep{SLN93}. Follow up studies have confirmed this 
detection and have also shown the presence of the isotope in BD+26$^o$~3578 
\citep{SLN98,HT97}. 
Unexpected $^6$Li abundances have been observed during a long flare on a
chromospherically active binary \citep{MR98}. These have been attributed to
spallation reactions. 

In this paper we present high resolution VLT UVES 
observations of the K star GJ~117 with emphasis on the Li 6707.8\AA\ 
line profile. GJ~117 (=HD~17925) is a single chromospherically active K dwarf. 
The object shows evidence for photometric variability with a period of 6.5 days \citep{Cu92} while 
the weak color variations show that the star is redder at minimum light.   
Spectroscopic observations in the Ca II lines also show strong evidence for 
variability at both 
short and long timescales \citep{B95}. 
Our observations and analysis are presented in \S\ 2. 
In \S\ 2.1 we present stellar models generated using the PHOENIX code 
for the temperature and gravity of GJ~117. We present
the results, derived lithium parameters,  
and $\frac{^{6}Li}{^{7}Li}$
ratio in \S\ 2.2.  Discussion of the results and implications for
lithium production are presented in \S\ 2.3, and lastly we summarize our
findings in \S\ 3.

\section{Observations \& Data Analysis}
\label{obs}
Measurements of $^6$Li in stellar spectra are very difficult and rely on the 
modeling of 
the increased width and red asymmetry of the 6707.8\AA\ doublet.  
The problem becomes particularly difficult in active late-type stars where 
large  rotational velocities will smear the line profiles. The problem is 
further complicated 
by the occurrence of blends, such as Ti~I and CN. 
The observations were conducted in December 2002 with the VLT Kueyen Telescope 
and UV-Visual 
Echelle Spectrograph (UVES).
This set-up used the 600 l~mm$^{-1}$ red grating and MIT/LL eev 
2k$\times$4k CCD. With an exposure time of 150 seconds this setup provided a resolution of $\approx$110,000 and a signal-to-noise of $\approx$400 
in the final spectrum. The VLT data were
reduced with the VLT data pipeline (uves/1.3.3) with further analysis carried out using the 
STARLINK based DIPSO software.
The UVES spectrum of GJ~117 in the Li~I 6708\AA\ region is shown in Figure 1.


\medskip
\subsection{Model Spectra}
\label{models}

Theoretical spectra were calculated using the general stellar atmosphere code
{\sc Phoenix} 
\citep{HS95, AH95}. 
We have used the LTE atmosphere models from the NextGen
series of \citep{haus99} for the effective temperature and gravity 
listed in Table~1. Given the uncertainties in T$_{eff}$ and
log($g$), typically $\pm$ 200K and $\pm$0.3 dex, we have used the directly 
converged 
models from NextGen which are sampled every 200K in
T$_{eff}$ and 0.5dex in log($g$). 
Line blanketing is handled within {\sc Phoenix} by means
of direct opacity sampling \citep{H01}.

The $^6$Li resonance doublet has been included in {\sc Phoenix} by adding the 
wavelengths and $gf$ values of \citet{SLN93} 
into the master line list \citet{K95} where the fraction of
$\frac{^{6}Li}{^{7}Li}$ is introduced as a parameter($\gamma$). In direct 
opacity sampling,
opacity at each wavelength point is calculated as a sum of opacities
from all the contributing species. The contributions from  $^7$Li and $^6$Li 
are then expressed as :

\begin{equation}
\kappa({^7Li})=(1-\gamma)~\kappa({^7Li_{tot}})
\end{equation}

\begin{equation}
\kappa({^6Li})=\gamma~\kappa({^6Li_{tot}})
\end{equation}
\noindent where $\kappa({^XLi_{tot}}$) is the opacity calculated as if all
lithium is in the X isotopic state. Theoretical line profiles have been 
calculated 
with $\gamma$ ranging from 0.0 to 0.2.
Microturbulence has been taken into account when generating the model spectra 
using a velocity of $\xi$=1.5 km s$^{-1}$. The $v$sini of the object was 
derived from $\chi^{2}$ analysis of several lines in the Li~I 6707.8\AA\ region.

\subsection{The $\frac{^{6}Li}{^{7}Li}$ isotope ratio}
We have carried out a comparison between the observed and model profiles using $\chi^{2}$ 
statistics. 
A grid of models in steps of $\frac{^{6}Li}{^{7}Li}$ = 0.01 was developed
and the most probable model with $\chi^{2}\sim$1 was determined. Figure~2 shows the 
$\Delta\chi^2$ computed for the model grid and plotted with the sign preserved.
The best fit has a $\frac{^{6}Li}{^{7}Li}$ = 0.030 $\pm$ 0.007 and $\Delta\chi^2$~$\sim$~30 as
compared to a model with no $^6$Li. 
The observed spectrum of GJ~117 fitted with $\frac{^{6}Li}{^{7}Li}$ 
ratios of 0.0, 0.03 and 0.10 is shown in Figure~3. 
$\chi^2$ was calculated for 35 degrees of freedom and 
a sign factor was added
to $\chi^2$ to indicate whether the observational data was larger than 
(positive) or smaller than (negative) the model.

The $\frac{^{6}Li}{^{7}Li}$ ratio derived from the best fits are also
given in Table~1.
1$\sigma$ errors for the table were derived using $\chi^2$ + 1 
e.g. \citet{N99}, and we also include a conservative 5\% systematic error
to account for the uncertainty in the continuum level making $\frac{^{6}Li}{^{7}Li}$ = 0.030$\pm$0.010. 

%
%

The influence of lines such as 
Fe I, Ti I, and CN on the Li I profile, can potentially mimic 
an enhanced $^6$Li abundance \citep{N99, R02}. The effects of possible CN bands
near 6707.5 have been seen to be $\leq$ 1 m\AA\ in the solar spectrum
and should have a negligible effect for GJ~117.
In earlier lower resolution data the Fe I line at 6707.4 \AA\
was often blended with Li I, however in the UVES observations
these are easily separable. We have tested models with enhanced Fe abundances
and found that these have no effect on the Li I line profile.

The most significant blend may come from the Ti I lines in the vicinity of 
6708\AA\ and could lead to erroneous results.  \citet{I01} have claimed the 
detection of $\frac{^{6}Li}{^{7}Li}$ = 0.12 in the  planet hosting star HD82943.
\citet{R02} concluded that this enhanced $^6$Li  could be explained by the
presence of a Ti I lines near 6708.03 and 6708.1 \AA. In a follow up paper  
\citet{I03} revised the ratio to $\frac{^{6}Li}{^{7}Li}$ = 0.05$\pm$0.02.  The
PHOENIX models used in our analysis were updated to include the atomic data for
Ti~I lines taken from Table~3 of \citet{R02}.  
We have investigated the
effects of Ti I on the Li line profile by constructing a grid of models with Ti
abundances ranging from log(Ti) 4.0 to 6.0 and fitting the Ti I line at
6716.6\AA. The best $\chi^2$ revealed an abundance of log(Ti) = 5.22 in
excellent agreement with the result of \citet{LH05}.  The latter value was used
in the determination of the $\frac{^{6}Li}{^{7}Li}$ ratio. The effects of Ti
abundance were investigated further by constructing an additional set of models
at a given Ti abundance with $\frac{^{6}Li}{^{7}Li}$ ranging from 0.0 to 0.1. The best fit to the Li line profile provided a
$\frac{^{6}Li}{^{7}Li}$ ratio at each Ti abundance. The results are shown in
Figure 4. The Ti abundance and $\frac{^{6}Li}{^{7}Li}$ ratio are anti-correlated
(also see \citet{R02}) in the sense that for a given Li profile, a lower Ti
abundance would lead to an increased $\frac{^{6}Li}{^{7}Li}$ ratio. However, in
the case of GJ 117 studied here, in order to drive the $\frac{^{6}Li}{^{7}Li}$
ratio to zero, a Ti abundance of five times greater (log(Ti) = 5.92) would be
required. We emphasize that both our results and those found in the literature 
are consistent with log(Ti) = 5.22. 

An additional check on any significant blending that may contaminate the red
wing of the lithium profile was done by comparing the spectrum of GJ 117 to the
template star HD~131511 which has a similar spectral type. 
The two objects have similar metallicity (Fe/H =0.13 to Fe/H = 0.15)
\citep{LH05} but HD~131511 has no detectable lithium \citep{T05}. HD 131511 was
observed at the McDonald 2.7m on 2005 May 24 with a resolution of $\approx$130,000 and a signal-to-noise just over 120. The spectra of the two objects are compared in
Fig 5.  Features in the 6707.8 to 6708.3 \AA\ region were examined on the HD~131511 spectrum and showed an equivalent width of less than 2 m\AA. We therefore conclude that our estimate of $\frac{^{6}Li}{^{7}Li}$ on GJ 117 is not effected significantly by other lines.

We have also analysed the bisectors of 4 nearby spectral lines of Fe~I  and 
Ca~I on  GJ~117 and found them to be vertical with an uncertainty of about 0.004
\AA\  (0.2 km/s). The red asymmetry in the Li  profile is therefore not caused
by the atmospheric effects of granulation. Our findings are consistent with
\citet{G05} who analyse the shape of the bisector of the K2V star $\epsilon$
Eri, a star with atmospheric properties similar to our object. Our findings are
also consistent with \citet{SLN98}. The lack of asymmetries can be explained by
the fact that  in active stars, like GJ~117, strong magnetic fields inhibit
convection which is known to be the case in sunspots.

\subsection{Can the $^6$Li abundance be attributed to enhanced activity ?}
\citet{DM95} have suggested that flare activity may account for   $^6$Li
production on HD 84937. However, \citet{Lem97}  have questioned the feasibility
of this scenario on the basis of the energetics  involved. We will now examine
if the activity of GJ~117 can account for the detected $^6$Li  on this source.
As already mentioned in the introduction GJ~117 has shown evidence for 
chromospheric activity. With a rotational velocity of 5.5 km
s$^{-1}$ GJ~117 is rotating  faster than the Sun and has a deeper convection
zone. Its X-ray luminosity is at least one  order of magnitude higher than the
X-ray luminosity of the Sun at solar maximum  \citep{MD89,HSS95}.  Its X-ray
emission implies a total average flare energy of  $\approx$
5~$\times$~10$^{29}$~ergs~s$^{-1}$ and a corresponding  total average U-band
flare energy of $\approx$~5~$\times$~10$^{28}$~ergs~s$^{-1}$ \citep{DB85,W95}. 
The most energetic solar flares are the X-class with energies of at least
10$^{32}$~ergs.  The GOES database shows that in 2000, a solar maximum year, 17
X-class flares were observed.  Assuming that these values also hold for GJ~117
they imply a total energy of E$\approx\,5\times10^{49}$~ergs over 1 Gyr. With a
production efficiency of 10$^{-3}$ atoms~ergs$^{-1}$ we have 5$\times
10^{46}$~atoms  \citep{L97}. The mass of the convection zone of GJ~117 is 0.06
M$_{\odot}$  \citep{Pi01}.  With an estimated log(Li) = 2.45 and 
$\frac{^{6}Li}{^{7}Li}$ = 0.03 we obtain $\approx$ 5$\times$10$^{44}$ atoms. In
the largest solar  flares the number of protons with energy over 30 MeV is more
than 10$^{33}$. This value implies  a 3$\times$10$^{29}$ $^7$Li atoms s$^{-1}$
\citep{L97} and 7$\times$10$^{28}$ $^6$Li  atoms s$^{-1}$ \citep{C80} or
2.2$\times$10$^{45}$ $^6$Li atoms in 1 Gyr.  Our estimates are therefore
consistent with the $^6$Li generation by spallation processes in flare events.

\section{Concluding Remarks}
\label{conc}
High resolution VLT UVES observations of the  dK1 star GJ~117 show the detection
of $^6$Li at the 3\% level ($\frac{^{6}Li}{^{7}Li}$ = 0.030$\pm0.010$).  Our
spectral analysis included the Ti I transitions and atomic data used by
\citet{R02} for similar studies of the planet hosting star HD~82493. GJ~117 is
considerably more active than the Sun.  We have examined the possibility of Li
production by spallation in stellar flares  and concluded that the energetics
involved are sufficient  for this process to take place.  Future work on this
subject could include the study of the Li profile during flare  events where an
enhanced $^6$Li may be expected.  

\acknowledgements
"Based on observations collected at the European Southern Observatory, Chile
(70.D-0075A)". We thank David Lambert for useful discussions and an anonymous
referee for suggested improvements.  PHH was supported in part by the P\^ole
Scientifique de Mod\'elisation Num\'erique at ENS-Lyon. Some of the calculations
presented here were performed at the H\"ochstleistungs Rechenzentrum Nord
(HLRN), at the National Energy Research Supercomputer Center (NERSC), supported
by the U.S. DOE, and at the San Diego Supercomputer Center (SDSC), supported by
the NSF. DC and MM are also grateful to the Defence Science and Technology Laboratory (dstl) for support under the Joint Grants Scheme.


\clearpage
\begin{figure}
\includegraphics[scale=0.65, angle=90]{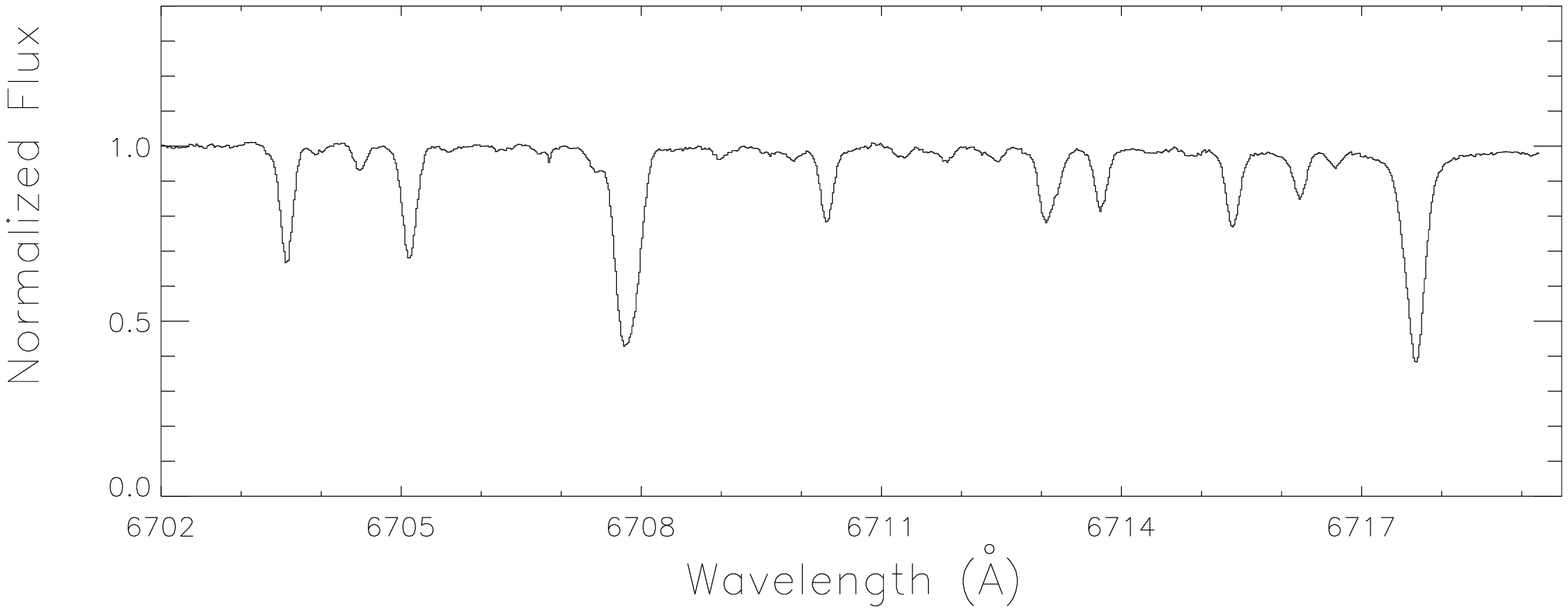}
\caption{
 VLT UVES spectra of GJ~117 showing the Li I 6707.8 \AA\ region. 
}
\end{figure}

\begin{figure}
\includegraphics[scale=0.70,angle=90]{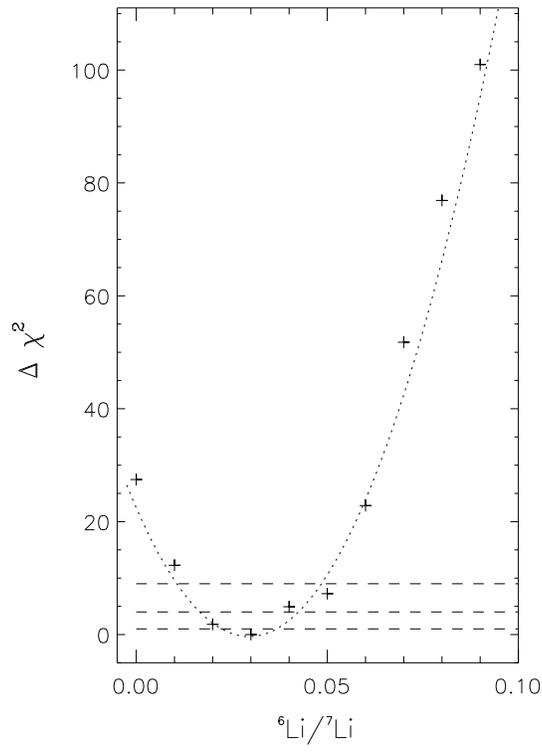}
\caption{
$\Delta\chi^2$ as a function of the $\frac{^{6}Li}{^{7}Li}$ ratio for Li I model line profiles 
for GJ~117 (see text). 
}
\end{figure}

\begin{figure}
\includegraphics[scale=0.65,angle=90]{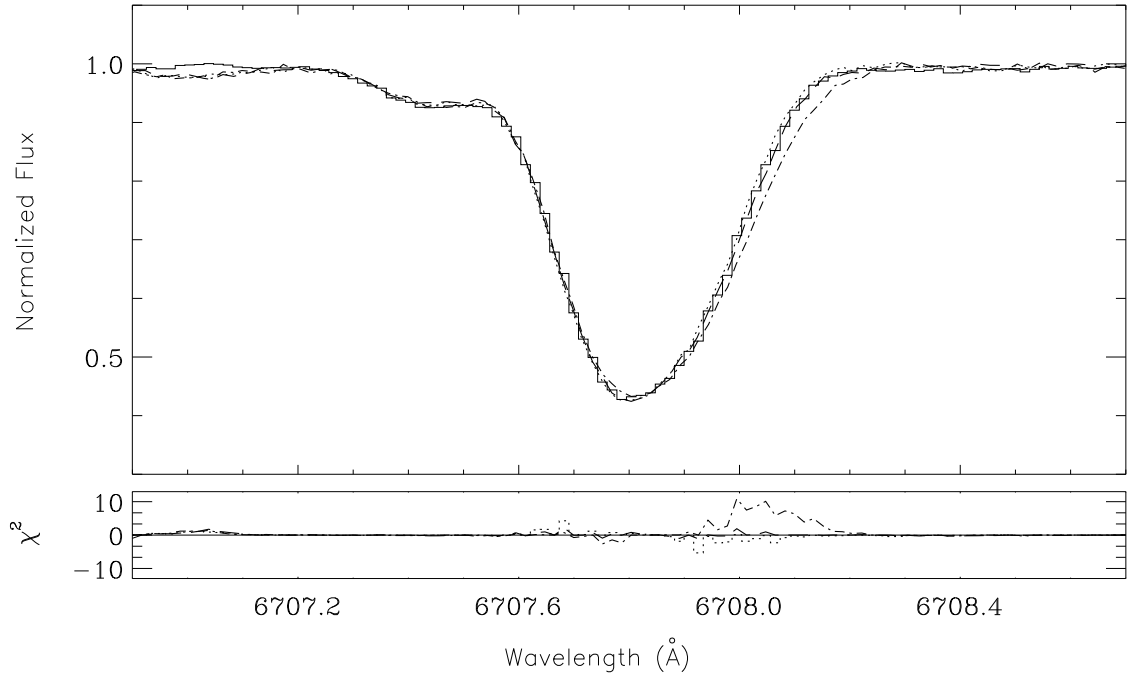}
\caption{
 GJ~117 spectrum for the Li I 6707.8 \AA\ region plotted as a solid histogram.
Over-plotted are PHOENIX models
with  $\frac{^{6}Li}{^{7}Li}$ ratios of 0.0 (dotted), 0.03 (dashed), 
and 0.10 (dash-dotted) lines. The lower panel shows $\chi^2$ from the difference
of the data minus the model (see text).
}

\end{figure}

\begin{figure}
\includegraphics[scale=0.65,angle=90]{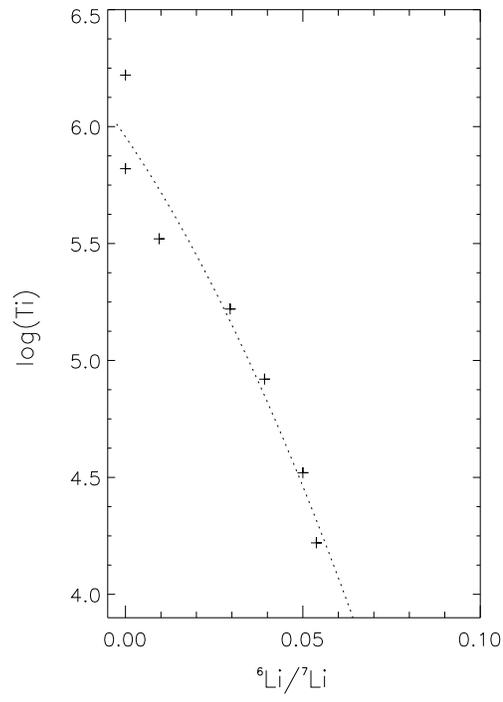}
\caption{
The Ti abundance as a function of the $\frac{^{6}Li}{^{7}Li}$ ratio
(see text).
}
\end{figure}

\begin{figure}
\includegraphics[scale=0.65,angle=90]{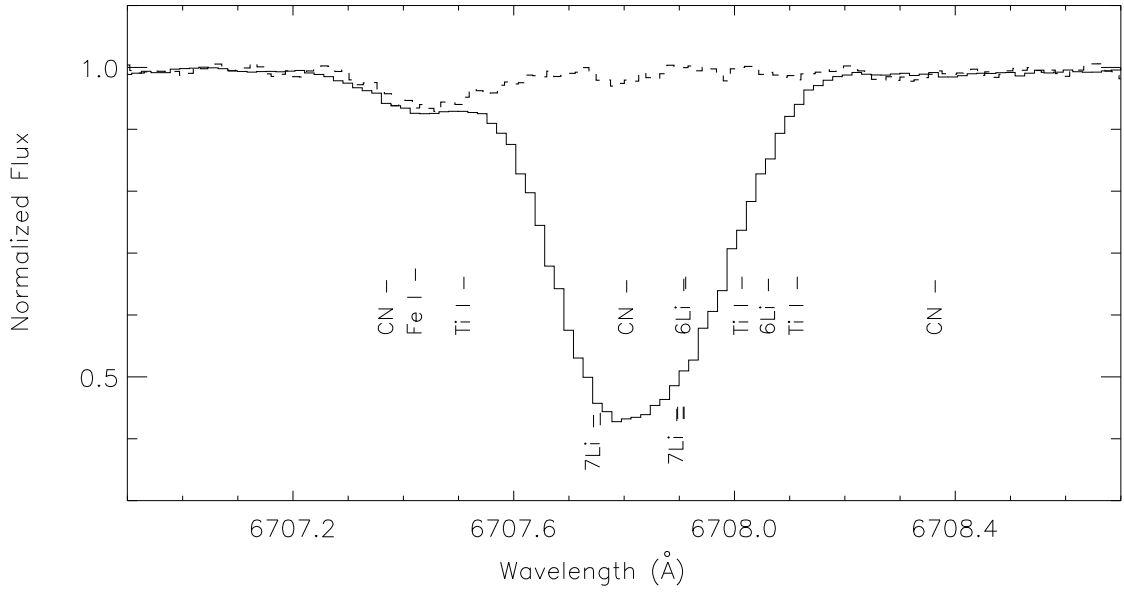}
\caption{
Comparison of the VLT GJ 117 spectrum (solid histogram) to that of a K2 template star, HD~131511 (dashed line).
HD~131511 was observed at the McDonald telescope in May 2005 (see text).
}

\end{figure}

\clearpage

\begin{deluxetable}{lcccccccr}
\tablewidth{0pt}
\tablenum{1}
\tablecaption{Stellar and Lithium Parameters}
\tablehead{
\colhead{Source}
  &\colhead{logT$_{eff}$}
  &\colhead{log $g$}
  &\colhead{$v$sini}
  &\colhead{Li eqw}
  &\colhead{log(Li)}
  &\colhead{$\frac{^{6}Li}{^{7}Li}$}
  &\colhead{}
\\
  \colhead{}
  &\colhead{}
  &\colhead{}
  &\colhead{ (km s$^{-1}$)}
  &\colhead{ (m\AA)}
  &\colhead{}
  &\colhead{}
  &\colhead{}
}
\startdata
GJ 117             & 3.70 & 4.6  & 5.5$\pm$0.2 & 211 & 2.45& 0.030$\pm$0.010    &     \\
\enddata
\end{deluxetable}


\begin{thebibliography}{}
\bibitem[Allard \& Hauschildt(1995)]{AH95}Allard, F.,Hauschildt, P.~H. 1995, \apj, 445, 433
\bibitem[Baliunas et al.(1995)]{B95} Baliunas,S.L. et al. 1995, \apj, 435, 269 
\bibitem[Canal et al.(1975)]{C75}Canal, R., Isern, J., \& Sanahuja, B. 1975, \apj, 200, 646
\bibitem[Canal et al.(1980)]{C80} Canal, R., Isern, J., Sanahuja, B. 1980, \apj, 235, 504
\bibitem[Cutispoto(1992)]{Cu92} Cutispoto, G. 1992, A\&AS, 95, 397 
\bibitem[Deliyannis \& Malaney(1995)]{DM95} Deliyannis, C.P., Malaney, R.A. 1995, ApJ, 453, 810
\bibitem[Doyle \& Butler(1985)]{DB85} Doyle, J.G., Butler, C.J. 1985, Nature, 313, 378
\bibitem[Gray(2005)]{G05}Gray, D.F. 2005, PASP, 117, 711
\bibitem[see Hauschildt et al.(1995)]{HS95}Hauschildt, P.~H., Starrfield, S., Allard, F., \& Baron,
  E. 1995, \apj, 447, 829
\bibitem[Hauschildt, Allard, \& Baron(1999)]{haus99} Hauschildt, P.~H., Allard, F., \& Baron, E.\ 1999, \apj, 512, 377 
\bibitem[Hauschildt et al.(2001)]{H01}Hauschildt, P. H., Lowenthal, David K., Baron, E. 2001, ApJS, 134, 323 
\bibitem[Hempelmann et al.(1995)]{HSS95} Hempelmann, A., Schmitt, J.H.M.M., Schultz, M., Ruediger, G., Stepien, K. 1995, \aap, 294, 515
\bibitem[Hobbs \& Thorburn(1997)]{HT97} Hobbs, L.~M.~\& Thorburn, J.~A.\ 1997, \apj, 491, 772 
\bibitem[Houdebine \& Doyle(1995)]{HD95} Houdebine, E.R., Doyle, J.G. 1995, \aap, 302, 861 
\bibitem[Giampapa(1984)]{G84} Giampapa, M., 1984 ApJ, 277, 235 
\bibitem[Israelian et al.(2001)]{I01}Israelian, G., Santos, N.C., Mayor, M, Rebolo, R. 2001, Nature, 411, 163
\bibitem[Israelian et al.(2003)]{I03} Israelian, G., Santos, N.C., Mayor, M., Rebolo, R. 2003, \aap, 405, 753 
\bibitem[Kurucz(1995)]{K95} Kurucz, R. L. 1995, in ASP Conf. Ser. 78, Astrophysical Applications of Powerful New Databases, ed. S. J. Adelman \& W. L. Wiese (San Francisco: ASP), 78, 205 
\bibitem[Lemoine et al.(1997)]{Lem97}Lemoine, M., Schramm, D.N., Truran, J.W., Coppi, C.J. 1997, \apj, 478, 554
\bibitem[Livshits(1997)]{L97}Livshits, M.~A. 1997, \solphys, 173, 377
\bibitem[Luck \& Heiter(2005)]{LH05}Luck, R.E., \& Heiter, U. 2005, \aj, 129, 1063
\bibitem[Mathioudakis \& Doyle(1989)]{MD89} Mathioudakis, M., Doyle, J.G. 1989, \aap, 224, 179
\bibitem[Montes \& Ramsey(1998)]{MR98} Montes, D., Ramsey, L.W. 1998, A\&A, 340, L5
\bibitem[see Mullan \& Linsky(1999)]{ML99} Mullan, D.J., Linsky, J. 1999, ApJ, 502, 511
\bibitem[M\"uller et al.(1975)]{M75} M\"uller, E.A., Peytremann, E., de La Reza, R. 1975, \solphys, 41. 53
\bibitem[Nissen et al.(1999)]{N99} Nissen, P.E., Lambert, D.I,, Primas, F., \& Smith V.V. 1999, \aap, 348, 211 
\bibitem[Pallavicini et al.(1992)]{P92} Pallavicini, R., et al. 1992, A\&A, 253, 185
\bibitem[Pinsonneault(1997)]{P97} Pinsonneault, M.H. 1997, ARA\&A, 35, 557
\bibitem[Pinsonneault et al.(2001)]{Pi01} Pinsonneault, M. H., DePoy, D. L., Coffee, M. 2001, \apj, 556, L59 
\bibitem[Reeves(1993)]{R93}Reeves, H. 1993, \aap, 269, 166  
\bibitem[Reddy et al.(2002)]{R02}Reddy, B.E., Lambert, D.L., Laws, C., Gonzalez, G., \& Covery, K. 2002, \mnras, 335, 1005
\bibitem[Ritzenhoff et al.(1997)]{R97}Ritzenhoff, S., Schr\"oter, E.H., \& Schmidt, W. 1997, \aap, 328, 695 
\bibitem[Russell(1996)]{Ru96} Russell, S.C. 1996, \apj, 463, 593 
\bibitem[Smith et al.(1993)]{SLN93} Smith, V.V., Lambert, D.I., Nissen, P.E. 1993, ApJ, 408, 262
\bibitem[Smith et al.(1998)]{SLN98} Smith, V.V., Lambert, D.I., Nissen, P.E. 1998, ApJ, 506, 405
\bibitem[Soderblom et al.(1993)]{S93} Soderblom, D.R., et. al. 1993, AJ, 106, 1059
\bibitem[Takeda \& Kawanomoto(2005)]{T05} Takeda, Y., Kawanomoto, S. 2005, PASJ, 57,45
\bibitem[Traub \& Roesler(1971)]{TR71} Traub, W., Roesler, F.L. 1971, \apj, 163, 629
\bibitem[Walker et al.(1985)]{W85}Walker, T.~P., Mathews, G.~J., \& Viola, V.~F. 1985, \apj, 745, 751
\bibitem[Whitehouse(1985)]{W95} Whitehouse, D.R. 1985, \aap, 145, 449

\end{thebibliography}
\end{document}